\documentclass[fdp,a4paper,fleqn%
]{w-art}
\usepackage{times,cite,w-thm}
\theoremstyle{plain}

\theoremstyle{definition}

\usepackage[]{graphicx}

\newcommand{\cc}[2]{c{#1\atopwithdelims[]#2}}

\begin{document}
\DOIsuffix{theDOIsuffix}
\Volume{55}
\Month{01}
\Year{2007}
\pagespan{1}{}
\Receiveddate{XXXX}
\Reviseddate{XXXX}
\Accepteddate{XXXX}
\Dateposted{XXXX}



\title[Towards classification of $SO(10)$ heterotic string vacua]{Towards classification of $SO(10)$ heterotic string vacua}


\author[J. Rizos]{John Rizos\inst{1,}%
  \footnote{Corresponding author\quad E-mail:~\textsf{irizos@uoi.gr},
            Phone: +30\,2651\,008\,614,
            Fax: +30\,2651\,008\,698}}
\address[\inst{1}]{Theory Division, Physics Department, GR45110 University of Ioannina, Greece}
\begin{abstract}
We report some recent progress towards classification of phenomenologically appealing heterotic string models
in the Free Fermionic Formulation. We focus on a class of $Z_2\times{Z_2}$ models  with $SO(10)$ space-time gauge symmetry and study
their main phenomenological aspects. We further consider reducing the  gauge symmetry to the Pati--Salam gauge group
${SU(4)}\times{SU(2)}_L\times{SU(2)}_R$ and impose a series of phenomenological constraints including the
existence of gauge symmetry breaking Higgs particles and the elimination of exotic fractionally charged states.
\end{abstract}
\maketitle                   





\section{Introduction}

String theory as a unified theory of all interactions should reproduce the Standard Model (SM) of gauge interactions at low energies.
Concrete semi-realistic 4-dimensional models reproducing Standard Model features have been constructed
in the past, in various formulations, with heterotic string models being among the most promising ones.
However, the quest of the Standard Model, among the huge number of 4-dimensional string vacua, requires a systematic exploration of
their phenomenological properties \cite{tyiim,hetclass,ttc,ffm,exo}.

Recently, we have developed tools for the study of a plethora of heterotic superstring vacua \cite{ffm} in the Free Fermionic Formulation \cite{fff}.
We have focused on a set of approximately $10^{12}$ models whose gauge sector includes an $SO(10)$ group factor than can naturally
accommodate the SM or some GUT model that can reduce to the SM. We have fully classified these vacua with respect to their phenomenological properties, as the number of $SO(10)$ spinorial representations, that accommodate the fermion generations, and the number of vectorials, that provide the SM
breaking Higgs doublets. Furthermore, we have considered the class of vacua where  $SO(10)$  is broken to the Pati--Salam (PS) gauge symmetry \cite{ps}
${SU(4)}\times{SU(2)}_L\times{SU(2)}_R$ and examined their main phenomenological characteristics: the number of fermion generations, the number of PS breaking Higgs multiplets, the number of SM breaking Higgs doublets, the number of additional vectorlike quarks and leptons and the number of exotic states.

The existence of fractional charge exotics is a generic property of spectrum of string models.
 These exotics, if  massless at the string level,  could be considered as a signature of string theory, however it seems hard to reconcile their presence with:
(i)  stringent experimental limits on fractional charged matter,
 (ii) the standard cosmological scenario, where relics from their production in the early universe could still be present,
 as the lightest fractional charge particle is stable,
(iii) coupling unification, due to their contribution to hypercharge running, even if they obtain intermediate scale masses.

The analysis of this class of approximately $2\times10^{15}$ Pati--Salam heterotic superstring vacua has demonstrated some of their attractive phenomenological properties as the existence of 3-generation vacua with appropriate Higgs to break the PS symmetry and the SM symmetry (approximately $2:10^4$ models)
 and the existence of exophobic models whose massless spectrum is free from  fractional charge exotics,  (approximately $1:10^6$ models).
\section{$SO(10)$ models}
A string model in the Free Fermionic Formulation, is defined by a set of $n$ basis vectors
 $B=\left\{v_1,v_2,\dots,v_n\right\}$ and a set of $2^{n(n-1)/2}$ phases $\cc{v_i}{v_j}, i>j=1,\dots,n$
 satisfying certain consistency constraints\cite{fff}. We examine a class of $SO(10)$ models  generated by
12 basis vectors
\begin{eqnarray}
v_1=1&=&\{\psi^\mu,\
\chi^{1,\dots,6},y^{1,\dots,6}, \omega^{1,\dots,6}|\bar{y}^{1,\dots,6},\bar{\omega}^{1,\dots,6},
\bar{\eta}^{1,2,3},
\bar{\psi}^{1,\dots,5},\bar{\phi}^{1,\dots,8}\},\nonumber\\
v_2=S&=&\{\psi^\mu,\chi^{1,\dots,6}\},\nonumber\\
v_{2+i}=e_i&=&\{y^{i},\omega^{i}|\bar{y}^i,\bar{\omega}^i\}, \
i=1,\dots,6,\nonumber\\
v_{9}=b_1&=&\{\chi^{34},\chi^{56},y^{34},y^{56}|\bar{y}^{34},
\bar{y}^{56},\bar{\eta}^1,\bar{\psi}^{1,\dots,5}\},\label{basis}\\
v_{10}=b_2&=&\{\chi^{12},\chi^{56},y^{12},y^{56}|\bar{y}^{12},
\bar{y}^{56},\bar{\eta}^2,\bar{\psi}^{1,\dots,5}\},\nonumber\\
v_{11}=z_1&=&\{\bar{\phi}^{1,\dots,4}\},\nonumber\\
v_{12}=z_2&=&\{\bar{\phi}^{5,\dots,8}\},\nonumber
\end{eqnarray}
and a set of $2^{12(12-1)/2}$ independent phases $\cc{v_i}{v_j}=\pm1, i>j=1,\dots,12$. We employ here the usual notation where the
20 left fermions are noted as $\psi^\mu, \chi^{1,\dots,6},\,y^{1,\dots,6},\,\omega^{1,\dots,6}$ and the 44 right ones as
$\bar{y}^{1,\dots,6},\bar{\omega}^{1,\dots,6},\,\bar{\eta}^{1,2,3},\,\bar{\psi}^{1,\dots,5},\,\bar{\phi}^{1,\dots,8}$.
The vector $v_1=1$ is required for consistency, vector $v_2=S$ is necessary for space-time supersymmetry, the vectors $e_i, i = 1, \dots, 6$ generate all possible symmetric shifts of the six internal fermionized coordinates, while the $b_1$, $b_2$ vectors generate the $Z_2\times{Z_2}$ twists which break $N=4$ supersymmetry to $N=1$. $z_1$ and $z_2$ are employed for the reduction of the hidden gauge group.

\begin{figure}[h]
\centering
\includegraphics[width=75mm]{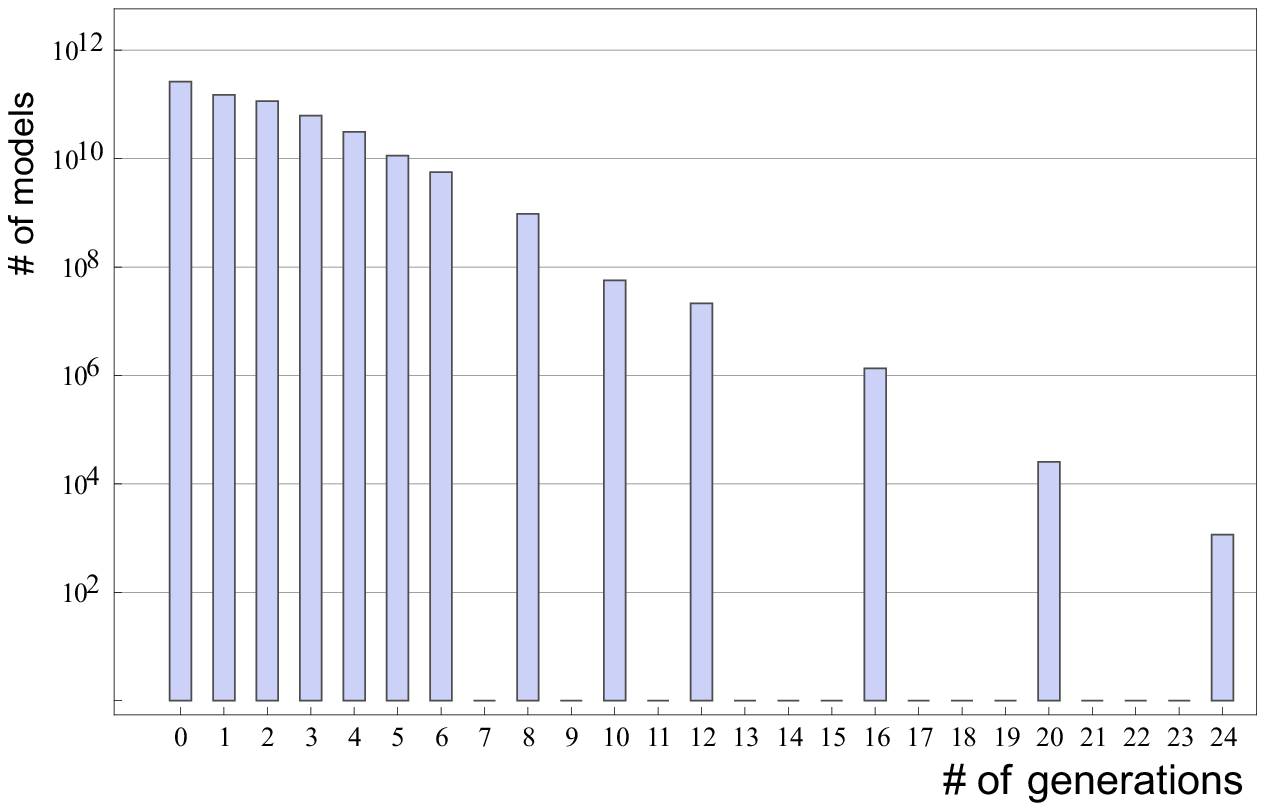}~a)\hfil
\includegraphics[width=50mm]{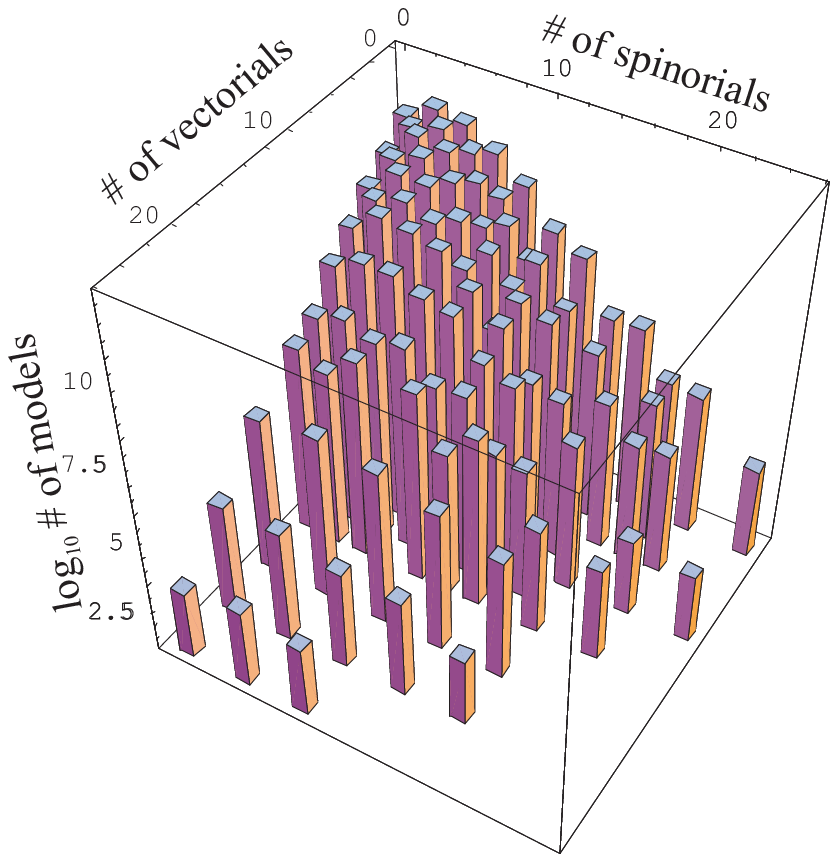}
~b)
\caption{Plot of the number of $SO(10)$ models as a function (a) of number of generations and (b) of the number of spinorial $\left(\mathbf{16}\,\text{and}\,\mathbf{\overline{16}}\right)$  and vectorial $\left(\mathbf{10}\right)$ twisted matter representations.}
\label{fig:svp}
\end{figure}

The full gauge group of the models generated by the basis  \eqref{basis}, is
${SO(10)}\times{U(1)}^3\times{SO(8)}^2$ (apart from certain enhancements).  $SO(10)$ matter spinor multiplets $\left(\mathbf{16},\mathbf{\overline{16}}\right)$ arise only from the twisted sectors
\begin{align}
B^1_{p_1 q_1 r_1 s_1} &=  b_1+p_1\,e_3+q_1\,e_4+r_1\,e_5+s_1\,e_6\nonumber\\
 & = \left\{x^{34},x^{56},(y_3\,\bar{y}_3)^{1-p_1},(y_4\,\bar{y}_4)^{1-q_1},
 (y_5\,\bar{y}_5)^{1-r_1},(y_6\,\bar{y}_6)^{1-s_1},\right.\nonumber\\
 &\ \ \ \ \ \ \ \ \left.(\omega_3\,\bar{\omega}_3)^{p_1},(\omega_4\,\bar{\omega}_4)^{q_1},
 (\omega_5\,\bar{\omega}_5)^{r_1},(\omega_6\,\bar{\omega}_6)^{s_1},\bar{\eta}^1,\bar{\psi}^{1,\dots,5}\right\}\\
B^2_{m_2 n_2 r_2 s_2} &= b_2+m_2\,e_1+n_2\,e_2+r_2\,e_5+s_2\,e_6\nonumber\\
 &= \left\{x^{12},x^{56},(y_1\,\bar{y}_1)^{1-m_2},(y_2\,\bar{y}_2)^{1-n_2},
 (y_5\,\bar{y}_5)^{1-r_2},(y_6\,\bar{y}_6)^{1-s_2},\right.\nonumber\\
 &\ \ \ \ \ \ \ \ \left.(\omega_1\,\bar{\omega}_2)^{m_2},(\omega_3\,\bar{\omega}_3)^{n_2},
 (\omega_5\,\bar{\omega}_5)^{r_2},(\omega_6\,\bar{\omega}_6)^{s_2},\bar{\eta}^2,\bar{\psi}^{1,\dots,5}\right\}\\
B^3_{m_3 n_3 p_3 q_3} &= b_3+m_3\,e_1+n_3\,e_2+p_3\,e_3+q_3\,e_4\nonumber\\
 & = \left\{x^{12},x^{34},(y_1\,\bar{y}_1)^{1-m_3},(y_2\,\bar{y}_2)^{1-n_3},
 (y_3\,\bar{y}_3)^{1-p_3},(y_4\,\bar{y}_4)^{1-q_3},\right.\nonumber\\
 &\ \ \ \ \ \ \ \ \left.(\omega_1\,\bar{\omega}_1)^{m_3},(\omega_2\,\bar{\omega}_2)^{n_3},
 (\omega_3\,\bar{\omega}_3)^{p_3},(\omega_4\,\bar{\omega}_4)^{q_3},\bar{\eta}^{3},\bar{\psi}^{1,\dots,5}\right\}
\end{align}
together with their supersymmetric partners from $S+B^1_{p_1,q_1,r_1,s_1},S+B^2_{m_2,n_2,r_2,s_2}$ and $S+B^3_{m_3,n_3,p_3,q_3}$,
where the upper index $I=1,2,3$ indicates the orbifold plane, $b_3=x+b_1+b_2$ and $m_i,n_i,p_i,q_i,r_i,s_i=0,1$.
Vectorial $SO(10)$ matter representations  $\left(\mathbf{10}\right)$ arise from   $V^I_{pqrs}=x+B^I_{pqrs},I=1,2,3$
 together with their superpartners from  $S+V^I_{pqrs}$.
Untwisted sector spectrum contains also 6 vectorial $SO(10)$ multiplets. Each twisted plane may produce up to $2^4=16$ spinor/antispinor multiplets and/or $2^4=16$ vector multiplets, however their multiplicities are reduced by projections imposed by generalized GSO conditions, which are determined from $\cc{v_i}{v_j}$.
After obtaining analytic formulae for the number of sninorials/antispinorials and vectorial matter representations and evaluating them using a fast
computer program we obtain a full classification of $SO(10)$ vacua. As shown in Figure \ref{fig:svp}a three generation models are quite abundant in this class (approximately 15\% of the models). The results for the number of models as a function of the total number of spinorials and
and vectorials are shown in Figure \ref{fig:svp}b. We conclude that models in this class appear in pairs symmetric under the exchange of the
number of spinorials with the number of vectorials, an interesting property known as ``spinor-vector duality" \cite{dual}. Many of the self-dual models, that is models with equal number of sninorials and vectorials appear often free of abelian anomalies, although the gauge symmetry does not enhance to $E_6$.

\section{Pati--Salam models}
The class of $SO(10)$ models under consideration possess interesting properties, however, they lack the necessary Higgs
particles that break $SO(10)$ to the SM. This is a not drawback of the specific class of  models but a consequence of the $k=1$ Kac-Moody algebra realization of  gauge symmetry on the string world-sheet \cite{frac}. However, the gauge symmetry breaking can be realized at the string level by introducing
additional basis vectors. In principle, flipped-$SU(5)$, Pati--Salam as well as the Standard Model gauge group can be obtained by this method.
The Pati--Salam gauge symmetry is the simplest to realize as it requires a single additional basis vector.

A heterotic string realization of a supersymmetric version of the  Pati--Salam (PS), $SU(4)\times{SU(2)}_L\times{SU(2)}_R$ , model has been studied in \cite{pss}. The quarks and leptons are accommodated in two  PS multiplets ${F}_L({\bf4},{\bf2}),{\bf1}+\bar{F}_R({\bf\bar 4},{\bf1},{\bf2})$
and the two pairs of SM breaking Higgs in the bi-doublet  $h({\bf1},{\bf2},{\bf2})$.
$D({\bf6},{\bf1},1)$ can accommodate additional d-quark like triplets. Using the standard hypercharge embedding, $Y=\frac{1}{\sqrt{6}}\,T_{15}+\frac{1}{2}\,I_{3L}+\frac{1}{2}\,I_{3R}$, that guarantees $sin^2\theta_w=\frac{3}{8}$
at the string scale,
states transforming as $({\bf1},{\bf1},{\bf2})$, $({\bf1},{\bf2},{\bf1})$, $({\bf4},{\bf1},{\bf1})$ accommodate fractional charge exotics.
The PS symmetry can be broken to the Standard Model by vevs to the neutral components  of
a pair of Higgs multiplets ${H}({\bf4},{\bf1},{\bf2}),\bar{H}({\bf\bar 4},{\bf1},{\bf2})$. The triplet remnants of the Higgs mechanism
could get heavy masses combined with the extra triplets in $D({\bf6},{\bf1},1)$.
Fermion masses are generated by the superpotential term
$
{F}_L({\bf4},{\bf2},{\bf1})\bar{F}_R({\bf\bar 4},{\bf1},{\bf2})\,h({\bf1},{\bf2},{\bf2})
$ while
neutrinos stay naturally light as they mix with additional heavy singlets \cite{pss}.

\begin{figure}
\centering
\includegraphics[width=80mm]{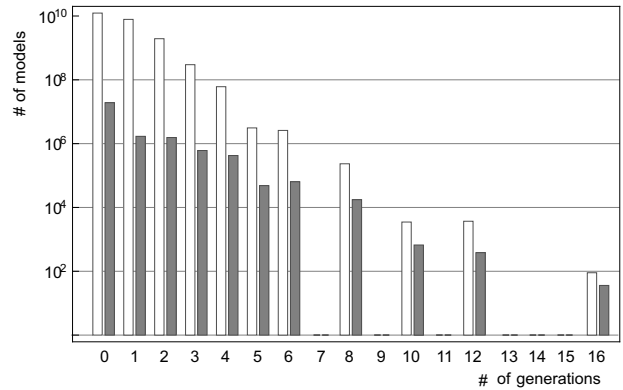}%
\caption{Total number of models (white) and number of exotic free models (gray) versus  number of generations in a random sample of $10^{11}$ Pati--Salam models. }
\label{fig:exo}
\end{figure}

Adding the vector
$
v_{13}=\alpha=\{\bar{\psi}^{45},\bar{\phi}^{1,2}\}
$
to the basis \eqref{basis} the full gauge group is broken down to $SU(4)\times{SU(2)}_L\times{SU(2)}_R\times{U(1)}^3\times{SU(2)}^4\times{SO(8)}$.
The 12 new  GSO projection phases $\cc{\alpha}{v_j},j=1,\dots,12$ increase the number of models to approximately $2\times10^{15}$. Following similar steps as in the $SO(10)$ case, we derive analytical formulae for the number of fermion generations, number of PS breaking Higgs multiplets,
number of SM breaking Higgs doublets, number of additional vectorlike quarks and leptons and number of exotic states. However, as  exploration
of all vacua in this class would require too much computer time, we prefer to analyze a subclass of randomly selected $10^{11}$ models.
We find some very interesting results: (i) the existence of 3-generation vacua with suitable Higgs to break the PS symmetry and the SM symmetry (approximately $2:10^4$ models)
(ii) the existence of exophobic models, that is a subclass of (i) whose spectrum is free from massless fractional charge exotics,  (approximately $1:10^6$ models).  As shown in Figure \ref{fig:exo}, exophobic models are not a
privilege of three generation models, they appear for all possible number of generations. The phenomenology of a specific PS model sharing all
 the above features is analyzed in \cite{newexo}.

\begin{acknowledgement}
This work is supported in part by the EU under contract PITN-GA-2009-237920.
\end{acknowledgement}

%
%

\end{document}